\documentclass[epj]{svjour}
\usepackage{latexsym}
\usepackage{graphics}
\usepackage{braket}
\usepackage{amsmath}
\usepackage{amssymb}
\usepackage{hyperref}
\usepackage{booktabs}
\usepackage{xcolor}
\newcommand{\acadd}[1]{{\color{blue}[Adrian: (add) \st{#1}]}}
\newcommand{\acrm}[1]{{\color{blue}[Adrian: (remove) \sout{#1}]}}

\begin{document}
\title{Upper limits on the temperature of inspiraling astrophysical black holes}
\author{Adrian Ka-Wai Chung 
\and Mairi Sakellariadou 
}                     
\institute{Theoretical Particle Physics and Cosmology Group, Department of Physics, King’s College London, University of London, Strand, London, WC2R 2LS, U.K.}
\date{Received: date / Revised version: date}
%
\abstract{
We present a method to constrain the temperature of astrophysical black holes through detecting the inspiral phase of binary black hole coalescences. 
At sufficient separation, inspiraling black holes can be regarded as isolated objects, hence their temperature can still be defined. 
Due to their intrinsic radiation, inspiraling black holes lose part of their masses during the inspiral phase. 
As a result, coalescence speeds up, introducing a  correction to the orbital phase. 
We show that this dephasing may allow us to constrain the temperature of inspiraling black holes through gravitational-wave detection.  
Using the binary black-hole coalescences of the first two observing runs of the Advanced LIGO and Virgo detectors, we constrain the temperature of parental black holes to be less than about $ 10^9 $ K. 
Such a constraint corresponds to luminosity of about $ 10^{-16} M_{\odot}  \rm s^{-1} $ for a black hole of $ 20 M_{\odot} $, which is  about 20 orders of magnitude below the peak luminosity of the corresponding gravitational-wave event, indicating no evidence for strong quantum-gravity effects through the detection of the inspiral phase.
\PACS{
      {PACS-key}{discribing text of that key}   \and
      {PACS-key}{discribing text of that key}
     } 
} 
\maketitle
\section{Introduction}
\label{intro}

The inspiral phase of a binary black hole system can be studied by analytical post-Newtonian calculations \cite{pN_01} \cite{pN_02} \cite{pN_03} \cite{pN_04} \cite{pN_05} \cite{pN_06}, or numerical relativity simulations \cite{Pretorius_01} \cite{Pretorius_02} \cite{Shapiro_01} \cite{Shibata_01}, considering the gravitational-wave luminosity (the binary’s
total energy flux at infinity) due to the relativistic corrections linked to the description of the source (multipole moments), and taking into account the binding energy of the system.
This is consistent with the detected gravitational-wave events \cite{LIGO_01} \cite{LIGO_02} \cite{LIGO_03} \cite{LIGO_04} \cite{LIGO_05} \cite{LIGO_06} \cite{LIGO_07} \cite{LIGO_08} \cite{LIGO_09} by the Advanced LIGO \cite{LIGO_detector_01} and Virgo detectors \cite{Virgo_detector_01}. 

However, multipole moments may not be the only contribution to the luminosity of binary black hole systems.
During the early inspiral phase, when black holes are sufficiently separated, they can be regarded as isolated objects, and hence can radiate \cite{BH_Radiation_01} \cite{BH_Radiation_02}. 
Thus, intrinsic black hole radiation may also contribute to the total gravitational-wave luminosity and affect the behaviour of binary black hole (BBH) coalescence.  
Note that particle creation of the surrounding spacetime, is negligible, hence not considered. 

In what follows, we study the effects of intrinsic black hole radiation on a BBH coalescence by means of post-Newtonian (PN) calculations. 
Although there are alternative gravitational theories which also predict possible reduction of individual masses during the inspiral phase, such as the scalar-tensor-vector theory \cite{SVTGravity_01} \cite{SVTGravity_02}, a maximal-reach-analysis can still be done to estimate the upper bound of the temperature of the parental black holes by assuming that the intrinsic black-hole radiation is the only effect reducing black-hole masses during the inspiral phase of the detected events. 
Moreover, to thoroughly constraint the temperature, we regard the temperature of inspiraling black holes as a free parameter.
We find that the intrinsic black-hole radiation results to a correction to the orbital phase, opening up a possibility to probe black hole radiation through inspiral phase detection. 
We then constrain the temperature of the parental black holes of the detected BBH coalescences during the first and second observing runs of the Advanced LIGO and Virgo detectors. 
We round up with a short discussion on the implications of our findings. 
Throughout this analysis, unless explicitly stated otherwise, we use $ c = G = \hbar = 1 $. 

\section{Orbital phase of BBH coalescence 
}
\label{sec:phase}

The inspiraling phase of a BBH coalescence can be observed by detecting gravitational waves at spatial infinity. 
Gravitational waves from the inspiral phase of a compact binary coalescence (CBC) can be expressed, in the frequency domain $ \tilde{h}(f) $, as
\begin{equation}
\tilde{h}(f) = A(f) e^{i \Psi(f)}, 
\end{equation}
where $ A(f) $ is the amplitude function and $ \Psi(f) $ the orbital phase of the CBC; they are both functions of frequency $ f $. 

Using the PN expansion of the Einstein field equation, $\Psi(f)$ obeys the differential equation \cite{pN_01}:
\begin{equation}\label{eq:phase_evolution}
\frac{d \Psi}{d \nu} = - \frac{\nu^3 E'(\nu)}{\mathcal{F}(\nu)}, 
\end{equation}
where $ \nu = (\pi M \omega)^{1/3} $ is the characteristic velocity of the binary with $ M $ the total mass of the system and $ \omega$ the instantaneous orbital frequency, $ E(\nu) $ is the binding energy (per unit mass) with a prime denoting its derivative with respect to the argument, and $ \mathcal{F}(\nu) $ is the energy flux. 

During the inspiraling phase of a BBH coalescence, $\nu\ll 1$, hence we can keep only the leading order term of the 3PN  expression for $ E(\nu) $ and the leading order of the 3.5PN expression  for $\mathcal{F}(\nu)$. 
More precisely \cite{pN_01} \cite{pN_02}
\begin{equation}\label{PN}
E_3(\nu)\approx -{1\over 2} \eta \nu^2\ \ ; \ \ 
\mathcal{F}_{3.5}(\nu)\approx {32\over 5}\eta^2\nu^{10},
\end{equation}
where $\eta = M_1 M_2 / (M_1 + M_2)^2 $ denotes the symmetric mass ratio.

During the inspiral phase, when the parental black holes are at sufficient separation, they can be considered as isolated objects emitting radiation \cite{BH_Radiation_01} \cite{BH_Radiation_02}.
Hence, they lose some of their mass and their horizon flux changes \cite{Horizon_flux_01} \cite{Horizon_Effect_01}.
We assume that the intrinsic black hole radiation is the only mechanism to reduce the black hole mass during the inspiral phase. 
Let us denote the luminosity of intrinsic radiation of each inspiraling black hole by $ \mathcal{F}_{\rm BHR, 1} $ and $ \mathcal{F}_{\rm BHR, 2} $, and the total luminosity by $ \mathcal{F}_{\rm BHR} = \mathcal{F}_{\rm BHR, 1} + \mathcal{F}_{\rm BHR, 2} $, which depends on the mass and temperature of the black holes (see Eq.~\ref{eq:F_BHR}).
The radiation power of black holes  in the source frame is given by the Stefan-Boltzmann law \cite{Don_Page_01}, 
\begin{equation}\label{eq:Page_power_01}
    \frac{d E}{d t} \approx \sigma_{B} A T^4,
\end{equation}
where $ \sigma_{B} \approx 5.67 \times 10^{-8} \rm W m^{-2} K^{-4}$ is the Stefan-Boltzmann constant, $A$ and $T$ are respectively the event-horizon area and temperature of the black hole. 
Hence, in the detector frame
\begin{equation} \label{eq:F_BHR}
\begin{split}
\mathcal{F}_{\rm BHR} & = \sigma_{B} (A_1 T_1^4 + A_2 T_2^4),  \\
A_{1, 2} & = 8 \pi M_{1, 2}^2 \left( 1 + \sqrt{1-\chi_{1, 2}^2}\right), 
\end{split}
\end{equation}
where $ M_{1, 2} $ are the red-shifted masses, $\chi_{1, 2}$ are dimensionless spins of the parental black holes and $ T_{1, 2} $ the corresponding red-shifted temperatures. 

Considering the intrinsic radiation to be small ($ \mathcal{F}_{\rm BHR} \ll \mathcal{F} $) and to remain small throughout the inspiral phase ($ \mathcal{F}_{\rm BHR} t \ll M $),
the total mass as a function of time, to leading order of $ \mathcal{F}_{\rm BHR} t $, reads
\begin{equation}\label{eq:change_01}
M(t) = M - \mathcal{F}_{\rm BHR} t, 
\end{equation}
where $ M $ is the initial total mass when the coordinate time $ t = 0 $. 
Similarly, $ \nu $ and $ \eta $ are also functions of time.
Since $ \mathcal{F}_{\rm BHR} \ll \mathcal{F} $ and $ \mathcal{F}_{\rm BHR} t \ll M $, we can write $ \nu(t) $ and $ \eta(t) $ up to the leading order in $ \mathcal{F}_{\rm BHR} $, 
\begin{equation}\label{eq:change_02}
\begin{split}
\nu(t) & \approx \nu - \frac{\nu}{3 M} \mathcal{F}_{\rm BHR} t; \\
\eta(t) & \approx \eta - \frac{M_1-M_2}{(M_1+M_2)^3} (\mathcal{F}_{\rm BHR, 2}M_1 - \mathcal{F}_{\rm BHR, 1} M_2)t, 
\end{split}
\end{equation}
where $ \nu $ and $ \eta $ are respectively the initial characteristic velocity of the binary and symmetric mass ratio at $ t = 0 $. 

The evolution of the total mass, Eq.~(\ref{eq:change_01}), would effect both the amplitude and the  phase of the emitted GWs.
As the Advanced LIGO and Virgo detectors are expected to be more sensitive to phases than  amplitudes \cite{LIGO_detector_01} \cite{Virgo_detector_01}, we shall focus on the effects on the former.
To calculate the change of phase, we first express the phase $\Psi(\nu)$ as
\begin{equation}\label{eq:phase}
\Psi(\nu) = \Psi^{(0)} (\nu) + \Psi_{\rm BHR} (\nu), 
\end{equation} 
where $ \Psi^{(0)} (\nu) $ is the orbital phase without considering the intrinsic black hole radiation and $ \Psi_{\rm BHR} (\nu) $ is the leading order correction due to black hole radiation, almost of the same order as $ \mathcal{F}_{\rm BHR} $.
Since $ \mathcal{F}_{\rm BHR} \ll \mathcal{F} $ and $ \mathcal{F}_{\rm BHR} t \ll M $, we can expand  $ \nu^3 E'(\nu) / \mathcal{F}(\nu) $, the RHS of Eq.~
(\ref{eq:phase_evolution}), up to leading order $ \mathcal{F}_{\rm BHR} t $.
Post-Newtonian calculations show that $ t $ can also be expressed in terms of $ \nu $. 
During the early inspiral phase when $ \nu \ll 1 $, one has \cite{pN_02}
\begin{equation}\label{eq:PN_t}
t_{3.5}(\nu) \approx t_{\rm ref} - \frac{5 M}{256 \eta \nu^8}, 
\end{equation}
where $ t_{\rm ref} $ stands for the reference time. 
Thus,
\begin{equation}\label{eq:BHR_dephasing}
\Psi_{\rm BHR} \approx \sigma_{B} \frac{N(\nu)}{D(\nu)},
\end{equation}
where
\begin{equation}
\begin{split}
 D(\nu) & = 2236416 \eta^3 \nu^{13} M^3, \\
N(\nu) & = 24875 A_1 M_1^3 T_1^4 + 24850 \eta A_1 M_1^3 T_1^4 - 525 M A_1 M_1 M_2 T_1^4 \\
& + 74625 A_1 M_1^2 M_2 T_1^4 + 74550 \eta A_1 M_1^2 M_2 T_1^4  \\
& + 525 M A_1 M_2^2 T_1^4 + 74625 A_1 M_1 M_2^2 T_1^4 \\
& + 74550 \eta A_1 M_1 M_2^2 T_1^4 + 24875 A_1 M_2^3 T_1^4 \\
& + 24850 \eta A_1 M_2^3 T_1^4 + 525 M A_2 M_1^2 T_2^4 \\
& + 24875 A_2 M_1^3 T_2^4 + 24850 \eta A_2 M_1^3 T_2^4 \\
& - 525 M A_2 M_1 M_2 T_2^4 + 74625 A_2 M_1^2 M_2 T_2^4 \\
& + 74550 \eta A_2 M_1^2 M_2 T_2^4 + 74625 A_2 M_1 M_2^2 T_2^4 \\
& + 74550 \eta A_2 M_1 M_2^2 T_2^4 + 24875 A_2 M_2^3 T_2^4 \\
& + 24850 \eta A_2 M_2^3 T_2^4. 
\end{split}
\nonumber
\end{equation}
There are a few points to note about $ \Psi_{\rm BHR} $. 
Firstly, $ \Psi_{\rm BHR} $ has opposite sign to the $ \nu $-dependent terms of $ \Psi^{(0)} $ (negative by convention), implying that coalescence will take place earlier than in the absence of black hole radiation. 
In other words, the intrinsic black hole radiation speeds up the coalescence process.
This can be explained by the fact that the BBH system loses energy more rapidly due to the intrinsic black hole radiation. 
Secondly, $ \Psi_{\rm BHR} $ is larger for smaller $ \nu $, suggesting that the intrinsic black hole radiation affects the inspiral more significantly during the early inspiral phase when gravitational-wave luminosity is of the similar order with the intrinsic black hole radiation. 
Thirdly, Eq.~(\ref{eq:BHR_dephasing}) is invariant if we exchange the labels ($ 1 \Longleftrightarrow 2 $), as expected, and its the right-hand side is dimensionless. 
Equations~(\ref{eq:F_BHR}), (\ref{eq:BHR_dephasing}) imply that the inspiral phase of binary black hole coalescences depends on the temperature,  opening up the possibility to constrain the black hole temperature through the signal of the inspiral phase.

\section{Parameter Estimation}
\label{sec:PE}

To constrain the temperature in a model-independent way, we consider $T_1$ and $T_2$ as free parameters independent of the black hole mass and spin. 
We also assume the Stefan-Boltzmann law \cite{Stefan} \cite{Boltzmann} to be valid for astrophysical black holes even if their temperature is not the Hawking temperature.

\subsection{Methodology}
\label{sec:Set_up}

We estimate $ T_{1} $, $T_2$ and the parameters of the source BBH $\vec{\theta}$ \footnote{From now on we denote the more massive black hole by the subscript ``1'' and the less massive one by  ``2''. } (e.g. masses, spins, luminosity distance and sky location etc) for the detected BBH using \texttt{Bilby} \cite{bilby}. 
We implement the  correction of orbital phase into the \texttt{IMRPhenomPv2} waveform model \cite{IMRPhenom_01} \cite{IMRPhenom_02}, \cite{IMRPhenom_03}.
Specifically, we estimate the posterior of the base-10 log of $ T_{1, 2}$ (denoted as $ \log_{10} T_{1, 2} $) given measured strain data $ \tilde{d} $ using Bayes' theorem,
\begin{equation}\label{eq:Bayes_theorem}
\begin{split}
& p(\log_{10} T_{1}, \log_{10} T_{2}, \vec{\theta} |\tilde{d}, \mathcal{H}_1, I) \\ 
& \propto p(\tilde{d}|\log_{10} T_{1}, \log_{10} T_{2}, \vec{\theta}, \mathcal{H}_1, I) \\
& \quad \times p(\log_{10} T_{1}, \log_{10} T_{2}, \vec{\theta}| \mathcal{H}_1, I);
\end{split}
\end{equation}
$\mathcal{H}_1 $ stands for our hypothesis that the parental black holes are emitting intrinsic radiation at temperatures $T_{1} $ and $T_2 $,
$p(\tilde{d}|\log_{10} T_{1}, \log_{10} T_{2}, \vec{\theta}, \mathcal{H}_1, I) $ is the likelihood function,
$p(\log_{10} T_{1} \log_{10} T_{2}, \vec{\theta}| \mathcal{H}_1, I) $ is the prior probability of $\log_{\rm 10} T_{1, 2} $ and $\vec{\theta}$. Note that $ T_{1, 2} $ are measured in Kelvin. 

Specifically, we prescribe a uniform prior for  $\log_{10} T_{1, 2} \in [0, 12]$. 
The lower limit of the prior corresponds to the order of magnitude of the cosmological microwave background, which should be the lower limit of the temperature measurement of all celestial objects in the Universe. 
We extend the upper limit to a value corresponding to the radiation power of about $ 10^2 M_{\odot} \rm s^{-1} $ for a black hole of mass $ 10 M_{\odot}$, which is the typical order of magnitude of the peak luminosity of gravitational-wave events. 
Note that the Hawking temperature $\sim 10^{-9} \rm K $ is not within in our prior. 

The likelihood of $ p(\tilde{d}|\log_{10} T_{1}, \log_{10} T_{2}, \vec{\theta}, \mathcal{H}_1, I) $ is \cite{PE_01}
\begin{equation}
\begin{split}
& p(\tilde{d}|\log_{10} T_{1}, \log_{10} T_{2}, \vec{\theta}, \mathcal{H}_1, I) \\
& \propto \exp \left( - \frac{1}{2} \sum_{{\rm D} = {\rm H, L, V} } \braket{\tilde{h}_{\rm D}-\tilde{d}_{\rm D}|\tilde{h}_{\rm D}-\tilde{d}_{\rm D}} \right);
\end{split}
\end{equation}
tilde is the Fourier transform, $ \tilde{h}_{\rm D} = \tilde{h}_{\rm D} (\log_{10} T_{1}, \log_{10} T_{2}, \vec{\theta}) $ is the waveform generated by the modified waveform given $ \log_{10} T_{1, 2} $ and $\vec{\theta}$, and by $\langle a|b \rangle$ we denote the noise-weighted inner product \cite{Finn_01}, 
\begin{equation}
\Braket{a|b} = 4 \text{Re} \int_{0}^{+ \infty} \frac{\tilde{a}(f) \tilde{b}^{\dagger} (f)}{S_h (f)} \text{d} f; 
\end{equation}
with $ S_h (f) $ the power-spectral density of a given detector D (H, L,V  for Hanford, Livingston and Virgo, respectively). 

\subsection{GW150914}
\label{sec:GW150914}

\begin{figure}
\resizebox{0.5\textwidth}{!}{%
  \includegraphics{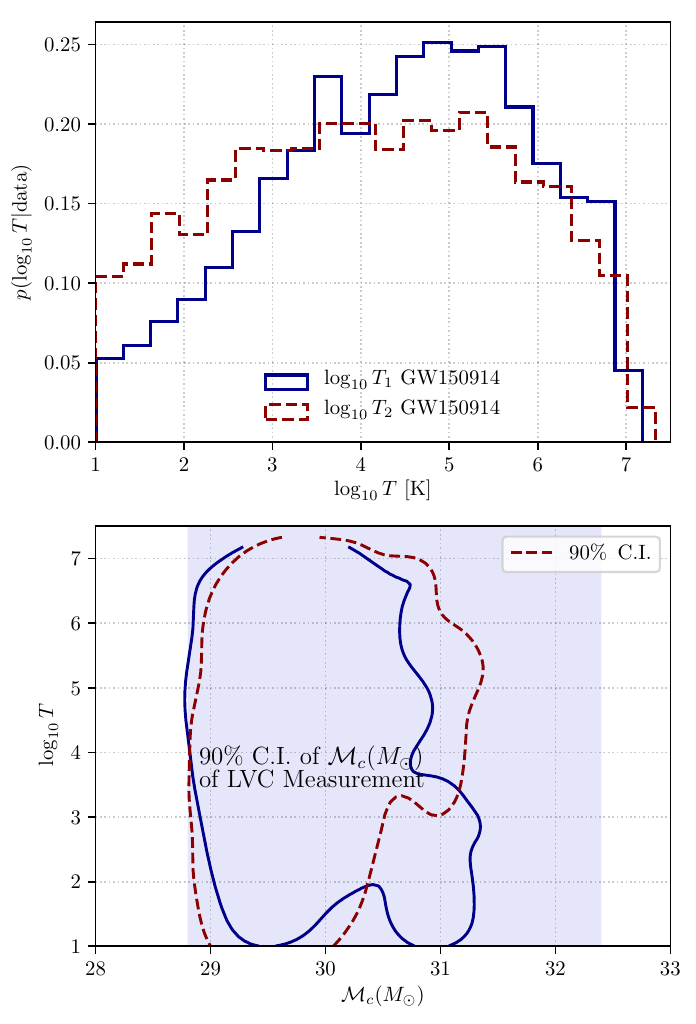}
}
\caption{Top panel: The marginalized posterior of $ \log_{10} T_{1} $ (solid blue line) and $ \log_{10} T_{2} $ (dashed red line) of the parental black holes of GW150914. 
The posteriors are of step-function shape as the detection of inspiral phase of GW150914 excludes completely the possibility of $ T_{\rm eff} > 10^7 \rm K$. 
Bottom panel: 90$\%$ confidence contour of the two-dimensional posterior of $ \log_{10} T_{1}$ (solid blue line) and $\log_{10} T_2 $ (dashed red line) vs. $\mathcal{M}_{c}$, the chirp mass in the detector frame. 
The contours show that there are no strong correlations between $ \log_{10} T_{1}$ (or $ \log_{10} T_{2}$) and $\mathcal{M}_c$. 
The shaded region shows the 90$\%$ percentile of $ \mathcal{M}_{c} $ of the first detection by Advanced LIGO and Virgo detectors \cite{LIGO_01}. 
}
\label{fig:Corner_Plot_GW150914}       
\end{figure}

We perform the analysis described in Sec.~\ref{sec:Set_up}  to obtain the posterior $ p(\log_{10} T_{1, 2}|{{\rm data}}) $ of $ \log_{10} T_{1} $ and $ \log_{10} T_{2} $ of the parental black holes of GW150914, the first detected gravitational wave event. 
The top panel of Fig.~\ref{fig:Corner_Plot_GW150914} shows the $ p(\log_{10} T_{1}|{{\rm data}}) $ (solid blue line) and $ p(\log_{10} T_{2}|{{\rm data}}) $ (dashed red line),  marginalized over other parameters, including the chirp mass, spins, calibration error of detectors and the temperature of the companion. 
The posteriors are sampled with 1024 live points. 
The posteriors of $\log_{10} T_{1, 2}$ have a in step-function shape around $ \sim 10^7 $ K. 
This result indicates that the intrinsic radiation power of GW150914's parental black holes is smaller than $10^{-17} M_{\odot} \rm s^{-1}$, which is $ \sim 10^{-19}$ of the peak luminosity. 
Hence, there is no evidence of intrinsic radiation from the progenitor black holes of GW150914. 

This constraint can be understood by a simple order-of-magnitude estimation. 
The parental masses of GW150914 are $M_1 \sim M_2 \sim 30 M_{\odot} $. 
Thus, $\eta \sim 0.25 $. 
For its inspiral phase, $ \nu = (\pi M \omega)^{1/3} \sim 10^{-1} $. 
Since the parental masses are of similar order of magnitudes, approximate $T_1 = T_2 = T_{\rm est} $. 
Thus, Eq.~(\ref{eq:BHR_dephasing}) implies
\begin{equation}
\begin{split}
\Psi_{\rm BHR} \sim 10^{-45}  T_{\rm est}^4.
\end{split}
\end{equation}
If we detect no signature of intrinsic radiation, $\Psi_{\rm BHR} \ll 1 $, implying 
$T_{\rm est} \ll 10^{10} \rm K $,  consistent with our result. 

The bottom panel of Fig.~\ref{fig:Corner_Plot_GW150914} shows the 90$\%$ confidence contour of the two-dimensional posterior of $ \log_{10} T_{1}$  (solid blue line) and $ \log_{10} T_{2}$  (dashed red line) vs. $\mathcal{M}_{c} $, where $ \mathcal{M}_c = (M_1 M_2)^{3/5} / (M_1 + M_2)^{1/5}$ is the chirp mass of GW150914 in the detector frame.
The contours show no strong correlation between $ \log_{10} T_{1} $ (or $ \log_{10} T_{2} $) and $ \mathcal{M}_c $. 
Therefore, $ \log_{10} T_{1} $ and $ \log_{10} T_{2} $ are not degenerate with $ \mathcal{M}_{c} $.
The shaded region shows the 90$\%$ percentile of $ \mathcal{M}_{c} $ of GW150914 estimated by the Advanced LIGO and Virgo detectors \cite{LIGO_01}.
The 90$\%$ confidence contour overlaps with the shaded region significantly, suggesting that the estimated $\mathcal{M}_{c} $ is consistent with those of known studies.
From Fig.~\ref{fig:Corner_Plot_GW150914}, we conclude that our test puts reasonable constraints on parental black holes' temperature while estimates $ \mathcal{M}_{c} $ accurately. 

\subsection{O1 and O2 Events}
\label{O1 and O2 Events}

\begin{figure}
\resizebox{0.5\textwidth}{!}{%
  \includegraphics{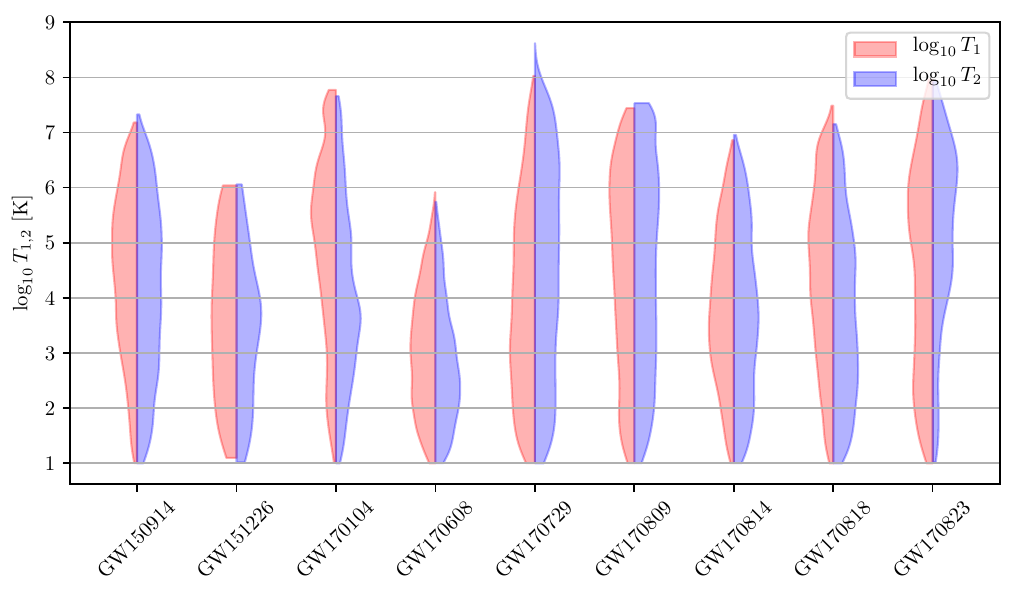}
}
\caption{Split violin plots which show the marginalized posteriors of $ \log_{10} T_{1} $ (left in red) and $\log_{10} T_2 $ (right in violet) of the parental black holes of each of the O1 and O2 (first and second observing run of the Advanced LIGO and Virgo) binary black-hole event (GW151012 is excluded from our analysis because it is not a confident detection). 
All posteriors exclude the possibility of $T_{1}$ and $ T_{2} > 10^9 \rm K$. 
Within the credible regions, the posteriors show no particularly strong support at any specific values of $T_1 $ and $T_2$, suggesting that we have no found any evidence of intrinsic radiation of the source binary black holes of O1 and O2 events. 
}
\label{fig:violin_plot}       
\end{figure}

We extend the analysis to all O1 and O2 binary black-hole events.
In Fig.~\ref{fig:violin_plot}, we plot the  posterior of $ \log_{10} T_{1} $ (red violin plots in left hand side) and $ \log_{10} T_{2} $ (violet violin plots in right hand side) for each detection.
For the O2 events which were detected after the Virgo detector had been online, its data were also included for the analysis. 
All posteriors show no support for of $T_1$ and $T_2 > 10^9 \rm K $. 
Most of the posteriors are of step-function shape, which are consistent with the results of GW150914 (see Fig.~\ref{fig:Corner_Plot_GW150914}). 
Similar to the case of GW150914, all posteriors show no particularly strong support at any specific values of $T_1$ and $T_2$. 
From Fig.~\ref{fig:violin_plot}, we conclude that the temperature of the parental black holes of all detected gravitational-wave events are $ \lessapprox 10^9 $ K and we have found no evidence of intrinsic radiations due to the progenitor black holes of O1 and O2 events. 

Our constraints of temperature correspond to extremely small black-hole radiation luminosity.
Table~\ref{tab:Temp} summarises the 90\% confidence interval of the posteriors of $ T_{1} $, $ T_{2} $ and the corresponding power of intrinsic black hole radiation, as computed using Eq.~(\ref{eq:F_BHR}). 
The 90\% percentiles correspond to black hole radiation powers which are much smaller than those of the peak luminosity of gravitational waves (about $ 10^2 M_{\odot} \rm s^{-1} $ for GW150914 \cite{LIGO_01}), consistent with our assumption of $ \mathcal{F}_{\rm BHR} \ll \mathcal{F} $ and keeping only the first terms in the expansion of $ \nu^3 E'(\nu) / \mathcal{F}(\nu) $. 
The constraints we have derived are  consistent with our assumption $ \mathcal{F}_{\rm BHR} t \ll M $. 
As shown in Table~\ref{tab:Temp}, for the analysed events $ \mathcal{F}_{\rm BHR} < 10^{-16} M_{\odot} \rm s^{-1}$ and the inspiral phase lasted for at most $ \sim 1 $s \cite{LIGO_04}. Hence, the mass loss of the parental black holes is less than $ 10^{-16} M_{\odot}$, which is much smaller than the smallest parental mass of the analysed events ($\sim 10 M_{\odot}$) \cite{LIGO_04}. 
Thus, Eqs.~(\ref{eq:change_01}) and (\ref{eq:BHR_dephasing}) are valid throughout the whole analysis. 
Since the constraint on the power of the intrinsic black hole radiation corresponds to only a tiny portion of the corresponding peak gravitational-wave luminosity, our results imply that we did not detect any evidence for abnormally strong black-hole radiations. 

\section{Concluding remarks}

We have presented a method to constrain the temperature of astrophysical black holes through the inspiral phase detection, without assuming a specific dependence of the temperature on either mass or spin.  
The masses of inspiraling black holes are reducing during the inspiral phase due to black hole radiation, speeding up the coalescence process and introducing a correction to the orbital phase.

By parameterising the dephasing in terms of temperature, one can constrain the temperature of the parental black holes of the gravitational-wave events detected by the Advanced LIGO and Virgo detectors during the first and second observing runs, to be less than about $  10^9 $ K. 
Constraints of this order of magnitude correspond to black hole radiation of about $ 10^{-16} M_{\odot} \text{s}^{-1}$ by a black hole of about $ 10 M_{\odot}$, which is a tiny fraction of the peak luminosity of the corresponding gravitational-wave event. 
Note that throughout the paper, we have assumed that the intrinsic black hole radiation is the only effect that can reduce the masses of inspiraling black holes during BBH coalescence. 
Our constraints can be further improved if more detail investigation of the tidal effects due to alternative gravitational effects \cite{Horizon_Effect_01} \cite{Horizon_flux_01} \cite{Tidal_heating_01} \cite{Tidal_heating_02}, are included. 

We find no evidence of intrinsic black hole radiation. 
This is expected because the predicted temperature of the progenitor black holes detected by the Advanced LIGO and Virgo detectors are $\sim 10^{-9} \rm K $, much lower than the cosmic microwave background temperature.
Unless there exist BBHs which consist of companion black holes of $ < 10^{-9} M_{\odot} $ (such that their temperature $> 3$K ), it is unlikely the intrinsic radiation will produce visible effects on the inspiral phase for our test to measure. 

Despite these difficulties, intrinsic black hole radiation by astrophysical black holes is worth to be investigated.
Although we focus on black hole thermodynamics in General Relativity, our analysis can also be adapted for black holes in alternative theories. 
In particular, the Hawking radiation increases significantly for black holes in higher-dimensional theories \cite{Hawking_radiation_higher_dim_01} \cite{Hawking_radiation_higher_dim_02} \cite{Hawking_radiation_higher_dim_03}, while it may not follow the usual Stefan-Boltzmann law \cite{Hawking_radiation_grey_body_01} \cite{Hawking_radiation_grey_body_02} \cite{Hawking_radiation_grey_body_03}.
Our analysis may lead to meaningful constraints higher-dimensional theories and provide a more thorough test of black-hole thermodynamics.

In the work we have presented here, we  considered a simple phenomenological approach, similar to  \cite{Horizon_Effect_01} \cite{Horizon_flux_01}, to derive the leading order term of the dephasing.
In a future study we plan to investigate how the emission of Hawking radiation can effect the dynamics of the BBH system in more detail.

\begin{table}
\caption{The 90\% of confidence interval (C.I.) of $ T_{\rm eff}$, in the detect frame, for the O1 and O2 binary black hole coalescence detected by Advanced LIGO and Virgo. 
The black hole radiation luminosity at the 90\% C.I. of $ T_{\rm eff} $ is also computed. 
The corresponding black hole radiation power is considerably smaller that the order of magnitude of the peak GW luminosity, suggesting no evidence of strong quantum gravity effects. }
\label{tab:Temp}       
\begin{tabular}{l@{\quad} c@{\quad} c@{\quad}}
\hline
\toprule
Event & 90 C.I. of $(T_{1}, T_{2})$ (in K)  & Power (in $ M_{\odot} s^{-1}$) \\ \hline
\midrule
GW150914                   &    $(2.05\times 10^6, 1.80\times 10^6)$    &  1.11 $ \times 10^{-18} $ \\
GW151226                   &    $(4.84\times 10^5, 1.55\times 10^5)$    &  3.60 $ \times 10^{-22} $ \\
GW170104                   &    $(1.18\times 10^7, 1.90\times 10^6)$    &  6.38 $ \times 10^{-16} $ \\
GW170608                   &    $(3.97\times 10^4, 2.84\times 10^4)$    &  1.17 $ \times 10^{-26} $ \\
GW170729                   &    $(2.36\times 10^6, 1.25\times 10^7)$    &  9.75 $ \times 10^{-16} $ \\ 
GW170809                   &    $(4.90\times 10^6, 1.21\times 10^7)$    &  4.40 $ \times 10^{-16} $ \\ 
GW170814                   &    $(4.95\times 10^5, 7.02\times 10^5)$    &  7.30 $ \times 10^{-21} $ \\
GW170818                   &    $(3.24\times 10^6, 8.79\times 10^5)$    &  4.83 $ \times 10^{-18} $ \\
GW170823                   &    $(6.80\times 10^6, 1.07\times 10^7)$    &  5.05 $ \times 10^{-16} $ \\ \hline
\end{tabular}
\vspace*{5cm}  
\end{table}
%
%
\section*{Acknowledgements}

We thank Gideon Koekoek and Nathan Johnson-McDaniel for reading carefully the manuscript and providing useful comments.
A.K.W.C. acknowledges Tjonnie Li and Ken K.Y. Ng for insightful discussions and Isaac C.F. Wong for advice in using \texttt{LALSuite}. 
A.K.W.C is supported by the Hong Kong Scholarship for Excellence Scheme (HKSES). 
The work of M.S. is supported in part by the Science and Technology Facility Council (STFC), United Kingdom, under the research grant ST/P000258/1. 

This research has made use of data, software and/or web tools obtained from the Gravitational Wave Open Science Center (https://www.gw-openscience.org), a service of LIGO Laboratory, the LIGO Scientific Collaboration and the Virgo Collaboration. LIGO is funded by the U.S. National Science Foundation. Virgo is funded by the French Centre National de Recherche Scientifique (CNRS), the Italian Istituto Nazionale della Fisica Nucleare (INFN) and the Dutch Nikhef, with contributions by Polish and Hungarian institutes.

This document carries a number of KCL-PH-TH/2020-07 and a LIGO document number LIGO-P2000068.

\end{document}